# ON THE POSSIBILITY OF USING DECAYLESS KINK OSCILLATIONS OF CORONAL LOOPS TO FORECAST POWERFUL SOLAR FLARES AND CORONAL MASS EJECTIONS

A.B. Nechaeva[1]*, I.V. Zimovets[1], I.N. Sharykin[1], S.A. Anfinogentov[1,2]

[1] *Space Research Institute of the RAS, Moscow, Russia*

[2] *Institute of Solar-Terrestrial Physics of the Siberian Branch of the RAS, Irkutsk, Russia*

This work investigates the decayless kink oscillations of solar coronal loops and examines possible changes in their behaviour in active regions (ARs) before powerful solar flares (M- and X-class) and in the absence of powerful flares. To this end, we analysed 14 ARs with powerful flares and 14 ARs without powerful flares. For each event, images obtained in the 171 Å and 94 Å AIA/SDO channels with 12-second cadence for 4 hours before the flare were retrieved and analysed. For ARs without powerful flares, arbitrary time intervals of similar duration were considered for comparison. Since the decayless oscillations have a very low amplitude (1–2 AIA/SDO pixels), we used the Motion Magnification technique to amplify the amplitude of these oscillations. Time-distance maps were constructed from the processed images in the 171 Å channel, from which oscillatory patterns were extracted 'manually'. Wavelet analysis was performed to check for changes in the oscillation period. No systematic changes were found. No obvious differences in the behaviour of oscillations in ARs with and without powerful flares were detected either. Additional information was obtained on coronal mass ejections (CMEs) from ARs in the vicinity of the time intervals under consideration. Based on the results of the analysis of a small sample of events, we came to the preliminary conclusion that the registration and analysis of decayless kink oscillations of high (~ 100–600 Mm) coronal loops based on this methodology is not promising for predicting powerful flares and CMEs.

## 1. INTRODUCTION

A solar flare is a sudden, rapid (~ 1–100 min) release of magnetic energy, characterised by an increase in brightness across a broad spectrum of electromagnetic radiation and observed in active regions (ARs) of the Sun [e.g., Benz, 2017]. Short-term forecasting of flares is an important applied task facing solar physics. Solar flares are preceded by its precursors. Precursors are, in a broad sense, a set of various phenomena in the

AR before (~ 1–2 h) the main flare [Martin, 1980; Wang et al., 2017]. They can manifest themselves as relatively weak short (seconds-minutes) bursts of radiation intensity from local areas of the AR, as well as in the form of increased oscillations or fluctuations in the X-ray and radio emission ranges [Kobrin et al., 1973; Zhdanov and Charikov, 1985; Chifor et al., 2007; Abramov-Maximov and Bakunina, 2020]. Apparently, the appearance of precursors is somehow related to the dynamics of magnetic fields and plasma in the AR during its evolution from a relatively quiet state to a flare state [Priest and Forbes, 2002; Wang et al., 2017; Toriumi and Wang, 2019]. However, the mechanisms of precursors and triggers of solar flares have not yet been clearly established, which hinders the development of methods for quantitative, physically based predictions of solar flares.

Coronal loops are one of the main structures of ARs. Observations show that coronal loops can undergo oscillations. In particular, there are two types of kink oscillations: decaying [Nakariakov et al., 1999; White and Verwichte, 2012; Goddard et al., 2016] and decayless [Wang et al., 2012; Anfinogentov et al., 2013; Nisticò et al., 2013]. The main features of the decayless mode are that the oscillations do not have a systematic trend of decreasing amplitude, which remains practically constant (with small variations) for a long time, on the order of several hours. Decayless oscillations are widespread in ARs and occur even in the absence of solar flares, eruptions, or other impulsive energy releases, unlike decaying oscillations. The periods of such oscillations are several minutes, and the amplitudes are on the order of several Mm or less. Although the mechanism of excitation of decayless oscillations is not yet clear [Nakariakov et al., 2016; Karampelas and Van Doorsselaere, 2020; Afanasyev et al., 2020], the period of decayless oscillations linearly correlates with the loop length, and therefore the oscillations are natural modes of the loop [Anfinogentov et al., 2015]. A review of the full spectrum of kink oscillations can be found in [Nakariakov et al., 2021]. The oscillations of solar coronal loops can be described in terms of the oscillations of a plasma cylinder in a magnetic field [Zajtsev and Stepanov, 1975; Roberts et al., 1984]. The period of the first harmonic (or fundamental period) of the kink oscillations of a plasma cylinder is expressed in terms of the loop length and kink velocity as follows:

$$P = 2L/C_K \qquad (1)$$

Here, $L$ is the length of the loop, and the kink velocity is $C_K = c_{Ai}(2\xi/(\xi + 1))^{1/2}$ where $\xi$ is the ratio of plasma densities outside and inside the loop, and the Alfvén velocity $c_{Ai}$ depends on the plasma density in the loop and the magnetic field. Based on this fact, it can be assumed that during the evolution of the parent AR, the period of kink oscillations of loops may change due to the restructuring of the magnetic field and plasma characteristics. The advantage of studying decayless oscillations of coronal loops in the context of flare precursors and the possibility of predicting them is that, firstly, decayless oscillations can be detected in many, if not all, coronal loops [Anfinogentov et al., 2015], and secondly, they can be observed for hours, i.e. on the timescales of AR preparation for flares. It is also very important that they are often clearly visible near the limb, where it is difficult to observe sunspots and measure magnetic fields on the photosphere, and to extrapolate them to the corona, which is used to make flare predictions [Bobra & Couvidat, 2015; Aschwanden, 2020; Zimovets & Sharykin, 2024]. Thus, our hypothesis is that when AR is preparing for a powerful flare, decayless oscillations of coronal loops may exhibit certain patterns in their evolution that could be used for forecasting, i.e., serve as special precursors to the flare.

To test this hypothesis, as a first step, we studied 14 ARs near the limb of the Sun where powerful M- and X-class flares occurred, as well as 14 ARs without powerful flares for comparison. For each AR, images were constructed using the Atmospheric Imaging Assembly (AIA) instrument aboard the Solar Dynamics Observatory (SDO) [Lemen et al., 2012] in the 171 Å (T ~ 0.6 MK) and 94 Å (T ~ 6 MK) channels. The time step was 12 seconds. The behaviour of loops in ARs was studied for time intervals of 4 hours preceding the flare, or for arbitrary time intervals of similar duration for ARs without powerful flares. We do not consider longer intervals due to the complexity of the methodology used. Below we describe this methodology and the results of its application for analysing decayless oscillations of coronal loops in the ARs on the Sun under consideration.

## 2. DATA ANALYSIS METHODOLOGY

To study the behaviour of decayless kink oscillations, 14 flares occurring on the solar limb were selected. The criteria for selecting flares were as follows: first, strong flares (above class M1.0) should be observed in the AR; second, coronal loops should be observed in the region, third, ARs near the limb (above 70 degrees longitude) were selected so that the loops could be observed with high contrast against the dark sky. The list of flares selected for analysis, indicating their position and class, is given in Table 1. In two cases, in addition to the X-class flare, there was also an M-class flare shortly before its onset, it is also included in the time interval under consideration (flares SOL2024-06-10T11:03 X1.6 and SOL2023-08-07T20:30 X1.5), which is also reflected in Table 1. The table also lists the parameters of coronal mass ejections (CMEs) associated with some of the flares under consideration, such as start time (ST), position angle (PA), angular width (AW) and linear speed (LS) from the CACTus [Robbrecht and Berghmans, 2004] and SOHO/LASCO CME Catalog (https://cdaw.gsfc.nasa.gov/CME_list/) [Gopalswamy et al., 2024]. The selection of associated CMEs was based on time (the start of the CME no later than an hour and a half after the flare) and position angle.

As already mentioned, AIA/SDO data in the 171 Å and 94 Å channels for 4 hours prior to the flare were used for analysis, with a 12-second interval. For each flare, AR images were downloaded in the 94 Å channel to construct a plot of the integral intensity of hot plasma and track AR changes, and in the 171 Å channel for further study of loop behaviour, since loops are best observed in this channel. Since the decayless oscillations have an amplitude of about the size of the AIA/SDO instrument pixel [Nisticò et al., 2013], the Motion Magnification algorithm [Anfinogentov and Nakariakov, 2016; Anfinogentov et al., 2019] was applied to the obtained AR images to amplify and better detect oscillations. The algorithm uses a dual tree complex wavelet transform (DTCWT) to amplify the amplitude of oscillatory processes whose time order is lower than $w$ by a factor of $k$. In this work, we use this algorithm with an amplification factor of $k = 5$ and $w = 80$ frames (16 minutes). Next, time-distance plots are constructed for the already amplified AR images using PyQtGraph (www.pyqtgraph.org). The sections for constructing the plots were selected manually based on how visually distinguishable the presence of an oscillatory pattern was for further analysis. The time-distance plots demonstrate the evolution of the intensity of the plasma's EUV radiation in the selected slice. Figure 1a shows an example of a time-distance plot for the SOL2023-08-07T20:30 flare. The upper right shows the AR in the 171 Å channel, the yellow circle shows the approximate location of the flare, and the yellow area shows the approximate area affected by the disturbance visible in the analysed AIA channels as a result of the flare, and the red line shows the slice that is selected for analysis. Fig. 1b shows another slice from the same AR for the same flare. Next, the observed oscillations are manually selected from the time-distance plots. The points selected at this stage are shown in blue in Figures 1a and 1b. Usually, the points are selected at the edge of the loop, where the contrast between the bright loop and the black background is greatest, which allows the oscillation pattern to be tracked well. Since the oscillations were selected manually and the points are unevenly distributed on the time grid, the resulting points were approximated with a smoothing spline, and the parasitic power trend was additionally subtracted. A wavelet analysis of the observed oscillations was performed on the data points taken from this curve. The resulting wavelet spectrograms for the full 4 hours prior to the flare are shown in Fig. 1a and 1b at the bottom left with the cone of influence, where edge effects are strong (translucent white), and the black curve shows the radiation intensity profile in the 94 Å channel. The dark grey contour indicates a 95% significance level. The 94 Å profile is also shown separately at the bottom right, where the vertical dotted line indicates the time of the main powerful flare, and the solid vertical grey lines indicate the times of the onset of weaker flares in this AR during the period under consideration. Similar results for two AR slices where the SOL2023-01-09T18:37 flare occurred are shown in Fig. 2. It should be noted that the spectrogram shows not the frequency, but the product of the frequency and the estimated loop length, since this value is directly proportional to the kink velocity according to (1). The loop length was estimated based on the assumption that the loop is a semicircle and that the loop length can be measured by knowing either the distance between the two bases (diameter) or between the centre point of the bases and the apex (radius). Errors arising from the projection of the loop onto the line of sight, possible changes in the shape of the loop during AR observation, and deviations of the loop shape from a semicircle are not taken into account due to their complexity. This process of estimating the loop length is similar to the process used in [Goddard et al., 2016; Nechaeva et al., 2019].

For comparison with decayless loop oscillations in ARs where no powerful flares (above class M1.0) occurred, 14 regions with well-observed coronal loops in the 171 Å channel were additionally analysed. The regions were selected so that no flares above class M1.0 occurred in them during the interval under consideration and several hours before and after it. A list of these regions with their coordinates and observation times is given in Table 2. Since there were no powerful flares in these ARs during the time intervals under consideration, there was no CME in most of the ARs under consideration, but for those cases where a CME was present, information about it is also included in the table. Spectrograms similar to those described earlier for such ARs are shown in Figures 3 and 4a.

To formalise the results of wavelet analysis of oscillations quantitatively, global oscillation spectra were also constructed, which represent the summation of wavelet spectrum values over time. Examples of such spectra can be seen in the upper panel of Fig. 5. The blue and red lines show the spectra for the two AR sections where the SOL2023-08-07T20:30 flare occurred, shown in Fig. 1. The black and green curves show the spectrum for oscillations from the same AR, but for the period from 16:00 to 20:00 UT on 2023-08-08, when there were no powerful flares in the region (Fig. 4). Each curve was approximated by a Gaussian (dotted line) to find the mean value of the frequency parameter multiplied by the loop length, and these mean values ⟨$v*L$⟩ and standard deviations are also listed in Tables 1 and 2. The tables also contain the estimated values of the length of the studied loop $L$ and the oscillation period $P$. The period was calculated from the values for $\langle v*L \rangle$ and $L$. It is important to note that several values are given for each flare, since several slices passing through different loops were analysed for one AR. In cases where spectrograms for the slice were obtained, but a specific peak could not be identified in the global spectrum, a dash (-) is indicated. For the global spectrum of the first slice for the SOL2023-08-07T20:30 flare (blue curve in Fig. 5), the average value is taken without taking into account the first peak in the spectrum, which we associate with parasitic periods. Also, for the third AR slice on 2023-08-08, whose oscillation spectrum is shown by the green curve, the average value $\langle v*L \rangle$ was not estimated due to the large number of peaks, which is reflected in Table 2 with the results. To compare the average values of the frequency multiplied by the loop length, we constructed histograms for cases of ARs with powerful flares and ARs without powerful flares, which are shown in the lower left panel of Fig. 5. Also, in Fig. 5, at the bottom centre and on the right, there are histograms for the estimated loop lengths and periods of decayless oscillations, respectively.

## 3. RESULTS AND CONCLUSIONS

When examining the evolution of decayless oscillations of 35 solar coronal loops during 4 hours prior to 14 powerful (M- and X-class) flares, no systematic changes in their behaviour were observed during the period under consideration. In some isolated cases, inhomogeneities in the behaviour of oscillations (e.g., the appearance of new frequencies or the amplification of existing ones) can be observed in the wavelet spectrograms, but no clear patterns in the appearance of these inhomogeneities were identified in this sample. These inhomogeneities may be related to physical manifestations of precursors in ARs (such as small flare events, mini-eruptions, and jets) or may be artefacts. This issue requires further investigation. In this work, we considered the change in the parameter frequency multiplied by the wavelength, i.e., the kink speed of oscillations with an accuracy of a constant according to formula (1). The evolution of other oscillation parameters was not considered, and it may also be of interest for further research. As a result of the comparison, no notable differences were found in the behaviour of loop oscillations before and after powerful flares. The average frequency multiplied by the loop length also does not differ for cases of ARs with and without powerful flares, as can be seen from the histograms in Fig. 5. The oscillation periods vary from a few minutes to about 20 minutes, but for large period values, the errors in their determination are also large. The period values obtained by a fairly rough estimate from global spectra (Fig. 5) coincide with those given in [Anfinogentov et al., 2013 and Anfinogentov et al., 2015]. Thus, we can make a preliminary conclusion that the pre-flare evolution of the AR does not significantly affect the kink velocity in the large loops (with loop lengths ~100–600 Mm) considered, which depends on the magnetic field strength in the loop, the plasma concentration, and the ratio of plasma densities inside and outside the loop. However, the study did not investigate shorter loops, which may be more sensitive to magnetic field restructuring in AR. In this work, we used data from the AIA/SDO instrument, whose resolution does not allow for the identification of short loops (with lengths less than 30 Mm), but there are a number of studies [e.g., Shrivastav et al., 2024; Shrivastav et al., 2025] that investigate decayless oscillations in low loops using data from the Extreme Ultraviolet Imager (EUI)/Solar Orbiter instrument, which has higher spatial and temporal resolution (~210 km/pixel at perihelion and 5 seconds, respectively). In the future, it may be interesting to study decayless oscillations in relatively shorter loops, which may be more sensitive to changes in the magnetic field and plasma concentration in the pre-flare phase of AR evolution.

In addition to examining the connection between decayless oscillations and flares, we also studied the connection with the presence of CMEs. For the ARs with powerful flares, CME was associated with the events in 12 out of 14 cases (86%), and in the case without flares, in 4 out of 14 cases (29%). It should also be noted that the 4 CME events found for the case without flares could be associated with both small class C flares in the ARs under consideration and have no relation to them at all, i.e., they could have occurred in other ARs. Since we did not find any obvious differences in the behaviour of decayless loop oscillations in 14 ARs with powerful flares and 14 ARs without powerful flares, we can make a preliminary conclusion that decayless oscillations of high (~100–600 Mm) loops practically 'do not respond' to the preparation of the ARs for the CME, as well as to powerful flares.

It is important to note that the data analysis method used in this work is not rigorous, since the sample of ARs for analysis is small (a total of 35 slices were analysed for situations where there was a powerful flare in the region, and 32 slices for situations where there was no flare), and a maximum of four slices were considered in each AR due to the quality of the source data, as we use data in a single narrow channel (171 Å). In some cases, it is impossible to track fluctuations on large time scales due to changes in the temperature of the loops, which become 'invisible' in the channel we have selected, but do not cease to exist. This problem can be solved by creating telescopes that observe coronal loops with higher angular resolution (~0.1" and above) simultaneously in a wider temperature range and by considering these data together. Limitations are also imposed by the manual selection of oscillations on time-distance plots for analysis and the visual comparison of the obtained wavelet spectrograms without the use of any strict statistical methods. Nevertheless, to our knowledge, this work is the first and original attempt to use decayless oscillations of coronal loops to predict powerful solar flares and CMEs, and therefore it may be of interest as a preliminary exploratory study.

## 4. ACKNOWLEDGEMENTS

We would like to thank the SDO/AIA team for providing free access to data, without which this work could not have been carried out. The SOHO/LASCO CME catalogue is generated and maintained at the CDAW Data Centre by NASA and The Catholic University of America in cooperation with the Naval Research Laboratory. SOHO is a project of international cooperation between ESA and NASA. We thank the reviewers for their helpful comments.

## FUNDING

This work was supported by a grant from the Russian Science Foundation (project No. 20-72-10158).

## CONFLICT OF INTEREST

The authors declare that they have no conflict of interest.

**Table 1.** List of flares used for analysis of decayless oscillations 4 hours before the flare, indicating their position and class, and associated coronal mass ejections with parameters: start time (ST), position angle (PA), angular width (AW), and linear speed (LS) from the CACTus and SOHO/LASCO CME Catalog catalogues. The last column shows the average value of the frequency*loop length $\langle v*L \rangle$ obtained from the global spectrum, as well as the loop length $L$ and oscillation period $P$ for each analysed slice.

| Flare | GOES Class // position of the AR (heliographic coordinates) | CME | $\langle v*L \rangle$, [Hz*Mm] $L$, [Mm] $P$, [min] |
|---|---|---|---|
| SOL2024-08-14T03:33 | M4.4 // S03W89 | ST 24/08/14 04:12 PA 276° (AW 212°) LS 546 km/s | 1.19 ± 0.43 336 4.71 ± 1.70 |
| | | | 2.68 ± 1.21 654 4.07 ± 1.83 |
| SOL2024-07-16T13:11 | X1.9 // S05W83 | ST 24/07/16 13:48 PA 277° (AW 92°) LS 520 ± 210 km/s | 0.33 ± 0.15 80 4.04 ± 1.84 |
| | | | 0.34 ± 0.29 116 5.69 ± 4.85 |
| | | | - 85 - |
| SOL2024-06-10T11:03 SOL2024-06-10T10:18 | X1.6 // S17W89 M5.3 // S17W89 | ST 24/06/10 10:48 PA 284° (AW 246°) LS 972 km/s | 0.83 ± 0.43 398 7.99 ± 4.14 |
| | | | 1.10 ± 0.32<br>450<br>6.82 ± 1.98 |
| | | | 0.81 ± 0.56<br>367<br>7.55 ± 5.22 |
| SOL2024-05-27T06:49 | X2.9 // S18E89 | ST 24/05/27 07:12<br>PA 75° (AW 228°)<br>LS 762 ± 319 km/s | 1.39 ± 0.49<br>596<br>7.15 ± 2.52 |
| | | | 1.89 ± 0.89<br>574<br>5.06 ± 2.38 |
| SOL2024-05-15T14:20 | X3.0 // S12E89 | - | 1.89 ± 0.68<br>610<br>5.38 ± 1.94 |
| | | | 1.82 ± 0.41<br>498<br>4.56 ± 1.02 |
| | | | 1.83 ± 0.34<br>498<br>4.53 ± 0.84 |
| SOL2024-05-15T08:15 | X3.5 // S18W89 | ST 24/05/15 08:48<br>PA 144° (AW 360°)<br>LS 1250 ± 685 km/s | -<br>671<br>- |
| | | | 1.37 ± 0.87<br>733<br>8.92 ± 5.66 |
| SOL2024-05-14T16:46 | X8.7 // S18W89 | ST 24/05/14 16:48<br>PA 244° (AW 34°)<br>LS 1037 ± 451 km/s | 1.14 ± 0.69<br>665<br>9.72 ± 5.79 |
| | | | 0.86 ± 0.55<br>629<br>12.19 ± 7.80 |
| SOL2024-05-14T12:40 | X1.2 // S17W89 | ST 24/05/14 13:00<br>PA 266° (AW 78°)<br>LS 749 km/s | -<br>590<br>- |
| | | | 0.52 ± 0.32<br>631<br>20.22 ± 12.44 |
| | | | -<br>229<br>- |
| SOL2024-02-09T12:53 | X3.4 // S37W85 | ST 24/02/09 13:25<br>PA 158° (AW 138°)<br>LS 992 ± 340 km/s | 1.19 ± 0.67<br>793<br>11.11 ± 6.25 |
| | | | 1.85 ± 1.61<br>635<br>5.72 ± 4.98 |
| | | | -<br>673<br>- |
| SOL2023-08-07T20:30<br>SOL2023-08-07T19:37 | X1.5 // N12W88<br>M1.4 // N19W75 | ST 23/08/07 20:36<br>PA 340° (AW 42°)<br>LS 416 ± 659 km/s | 0.37 ± 0.15<br>359<br>16.17 ± 6.56 |
| | | | 0.60 ± 0.11<br>274<br>7.61 ± 1.39 |
| | | | 0.45 ± 0.50<br>298<br>11.04 ± 12.26 |
| SOL2023-06-20T16:42 | X1.1 // S17E73 | ST 23/06/20 17:12<br>PA 87° (AW 108°)<br>LS 679 ± 81 km/s | 0.51 ± 0.53<br>207<br>6.76 ± 7.03 |
| SOL2023-01-09T18:37 | X1.9 // S13E71 | ST 23/01/09 19:12<br>PA 73° (AW 8°)<br>LS 618 ± 225 km/s | 0.80 ± 0.24<br>199<br>4.14 ± 1.24 |
| | | PA 46° (AW 14°)<br>LS 857 ± 87 km/s | 0.75 ± 0.41<br>301<br>6.69 ± 3.66 |
| | | | 0.89 ± 0.60<br>226<br>4.23 ± 2.85 |
| SOL2023-01-06T00:43 | X1.2 // S20E81 | - | -<br>261<br>- |
| SOL2022-04-30T13:37 | X1.1 // N15W100 | ST 22/04/30 14:00<br>PA 260° (AW 187°)<br>LS 498 km/s | 0.65 ± 0.10<br>149<br>3.82 ± 0.58 |
| | | | -<br>485<br>- |

| | | | 1.19 ± 0.83<br>564<br>7.90 ± 5.51 |
| --- | --- | --- | --- |
| | | | 1.26 ± 0.32<br>536<br>7.09 ± 1.80 |

**Table 2.** Active regions used for the analysis of decayless oscillations in the absence of powerful flares, the time interval during which they were considered, and their location. The last column shows the average value of the frequency*loop length ⟨υ*L⟩ obtained from the global spectrum, as well as the loop length *L* and oscillation period *P* for each analysed slice.

| Date and time interval UT | Position of the AR (heliographic coordinates) | CME | ⟨υ*L⟩, [Hz*Mm] L, [Mm] P,[min] |
|---|---|---|---|
| 2024-08-28 04:00-08:00 | S14E88 | - | 0.53 ± 0.17 232 7.29 ± 2.34 |
| | | | 0.89 ± 0.43 339 6.35 ± 3.07 |
| | | | 2.16 ± 0.63 601 4.64 ± 1.35 |
| 2024-07-18 04:00-08:00 | S9W86 | ST 24/07/18 07:24 PA 120° (AW 42°) LS 441 km/s | 1.44 ± 0.66 586 6.78 ± 3.11 |
| | | | - 612 - |
| 2024-06-11 08:00-12:00 | S13E86 | - | 1.21 ± 0.38 301 4.15 ± 1.40 |
| | | | 1.35 ± 0.52 354 4.37 ± 1.68 |

|  |  |  | 1.24 ± 1.06 554 7.45 ± 6.37 |
|---|---|---|---|
| 2024-05-19 04:00-08:00 | N21W89 | - | 1.12 ± 0.54 366 5.45 ± 2.62 |
|  |  |  | 1.34 ± 1.40 523 6.50 ± 6.79 |
| 2023-08-08 16:00 – 20:00 | N12W88 | - | 0.58 ± 0.28 230 6.61 ± 3.19 |
|  |  |  | 0.82 ± 0.20 266 5.41 ± 1.32 |
|  |  |  | - 104 - |
| 2023-07-20 08:00 – 12:00 | S25W85 | ST 23/07/20 09:48 PA 254° (AW 28°) LS 672 km/s | 1.55 ± 0.88 647 6.96 ± 3.95 |
|  |  |  | 1.51 ± 0.42 496 5.47 ± 1.52 |
| 2023-06-21 00:00 – 04:00 | N17E73 | - | - 226 - |
|  |  |  | 0.54 ± 0.21 |
| | | | 163<br>5.03 ± 1.96 |
| 2023-04-07<br>08:00 – 12:00 | N11W87 | - | 0.82 ± 0.24<br>220<br>4.47 ± 1.31 |
| | | | 1.42 ± 0.48<br>399<br>4.68 ± 1.58 |
| | | | 1.16 ± 0.89<br>397<br>5.70 ± 4.38 |
| 2023-04-01<br>16:00 – 20:00 | S22W84 | - | 1.64 ± 0.83<br>621<br>6.31 ± 3.19 |
| | | | 1.12 ± 0.85<br>390<br>5.80 ± 4.40 |
| | | | -<br>345<br>- |
| | | | 0.69 ± 0.47<br>308<br>7.43 ± 5.06 |
| | | | 0.78 ± 0.21<br>416<br>8.89 ± 2.39 |
| 2023-01-10<br>08:00 – 12:00 | S13E69 | ST 23/01/10 13:25<br>PA 75° (AW 32°) | 0.50 ± 0.27<br>162 |

|  |  |  | LS 473 km/s | 5.40 ± 2.91 |
| --- | --- | --- | --- | --- |
| 2023-01-07 08:00 – 12:00 | S13E80 |  | - | 0.26 ± 0.10 254 16.28 ± 6.26 |
| 2022-12-28 08:00 – 12:00 | S19E87 |  | - | 1.76 ± 0.82 612 5.79 ± 2.70 |
|  |  |  |  | 1.21 ± 0.25 303 4.17 ± 0.86 |
| 2022-12-18 08:00 – 12:00 | N25E88 |  | - | 1.01 ± 0.43 391 6.45 ± 2.75 |
| 2022-05-01 08:00 – 12:00 | N15W91 |  | ST 22/05/01 08:36 PA 251° (AW 240°) LS 700 km/s | 0.82 ± 0.26 398 8.09 ± 2.56 |
|  |  |  |  | 0.56 ± 0.41 365 10.86 ± 7.95 |

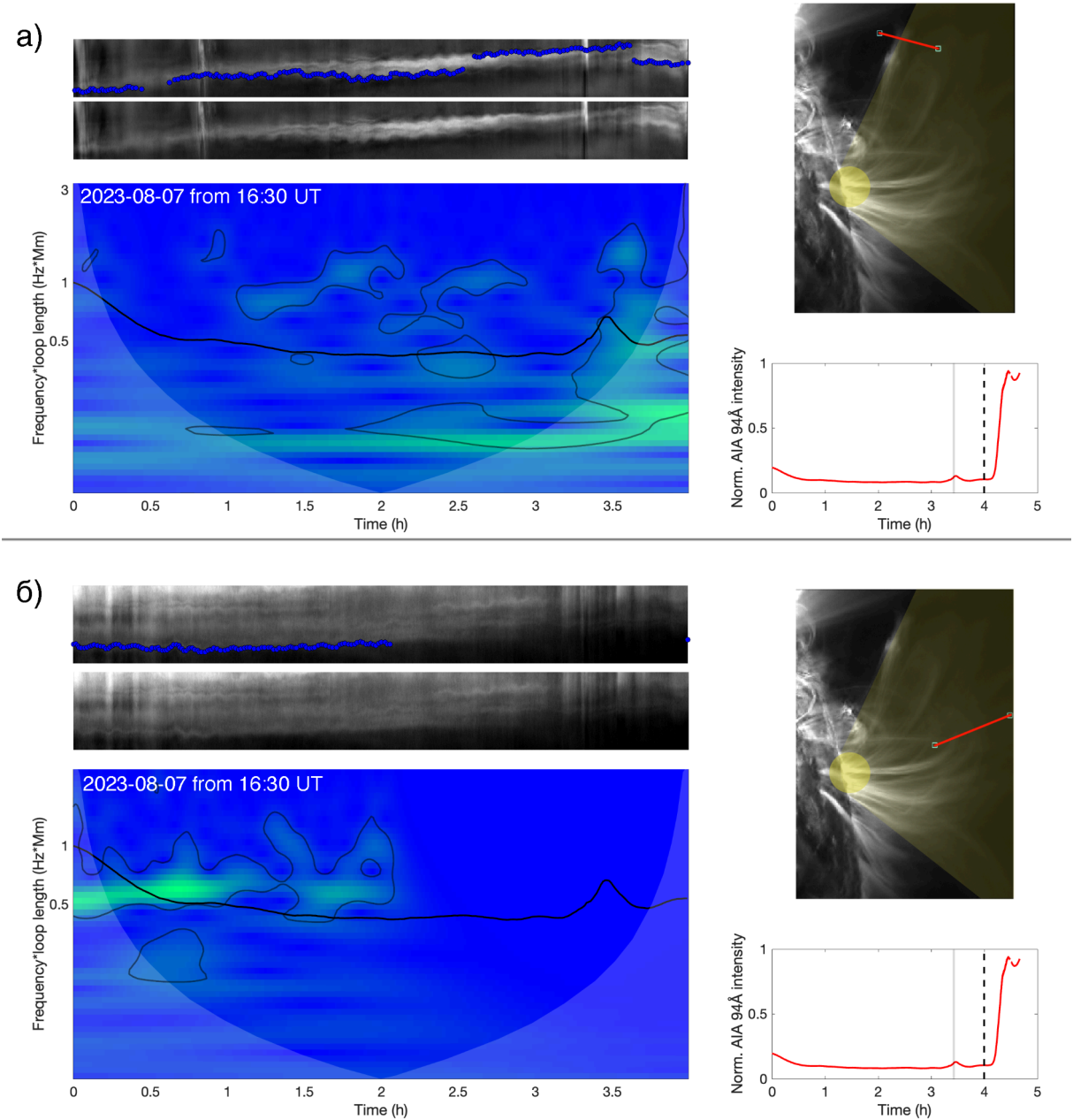

**Fig. 1.** Results of the analysis of two slices for the SOL2023-08-07T20:30 flare. Each of the panels a) and b) shows time-distance plots in the 171 Å channel with manually selected blue oscillation points for analysis. The flare onset is at *t*=4 h. Below is a wavelet spectrogram for the frequency*loop length parameter with a cone of influence, where edge effects are strong (translucent white). The black line shows the integral intensity curve of plasma radiation in the 94 Å channel from the AR under consideration. The upper right shows a representative image in the 171 Å channel of the AR itself with the studied section (red line). The approximate position of the flare and the cone affected by the disturbance are shown in yellow translucent. Below is the same intensity curve in the 94 Å channel with the start times of weaker flares marked with grey vertical lines: SOL2023-08-07T19:37 (M1.4).

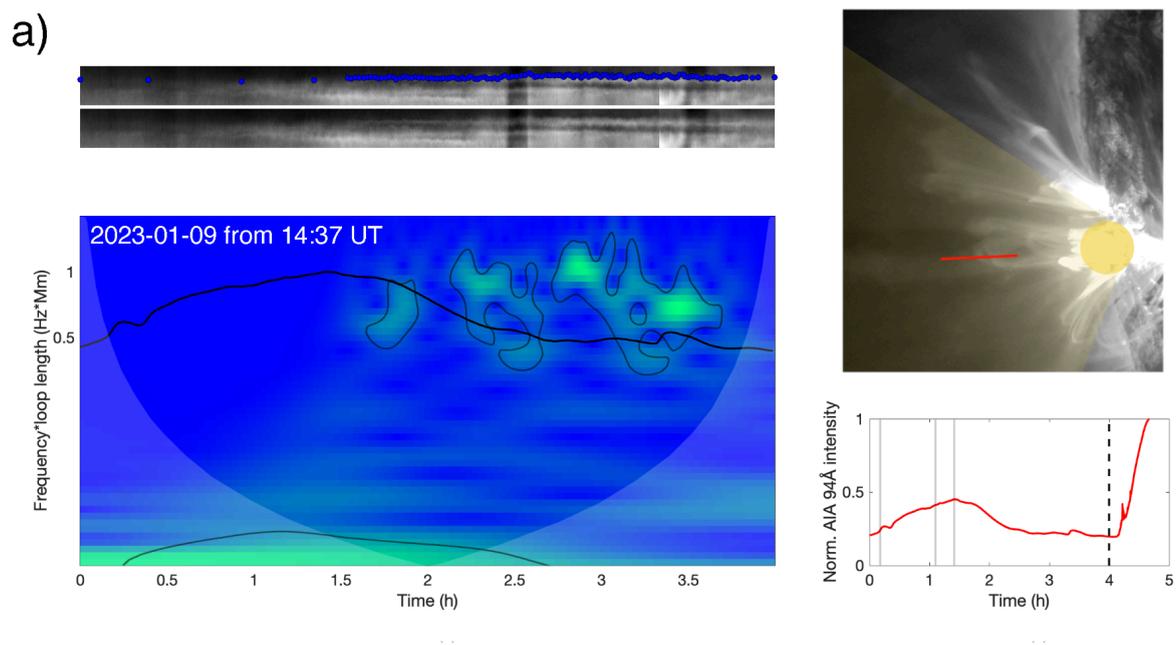

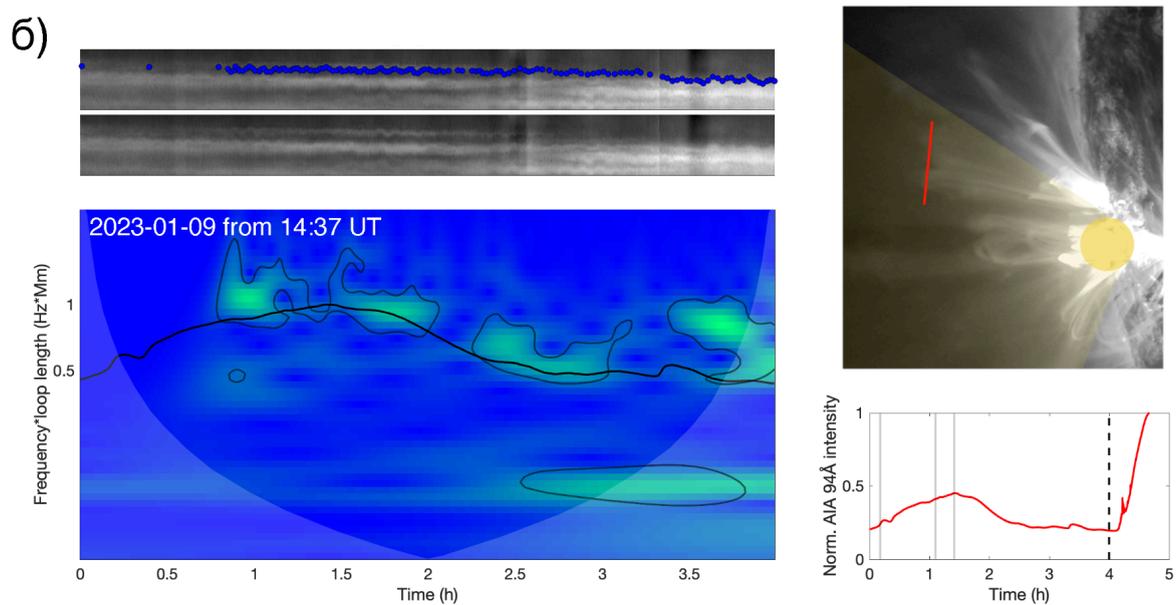

**Fig. 2.** Results of the analysis of two slices for the SOL2023-01-09T18:37 flare, similar to those shown in Fig. 1. Weaker flares in the region during the period under review: SOL2023-01-09T14:48 (C7.8), SOL2023-01-09T15:43 (C5.1), SOL2023-01-09T16:03 (C4.1).

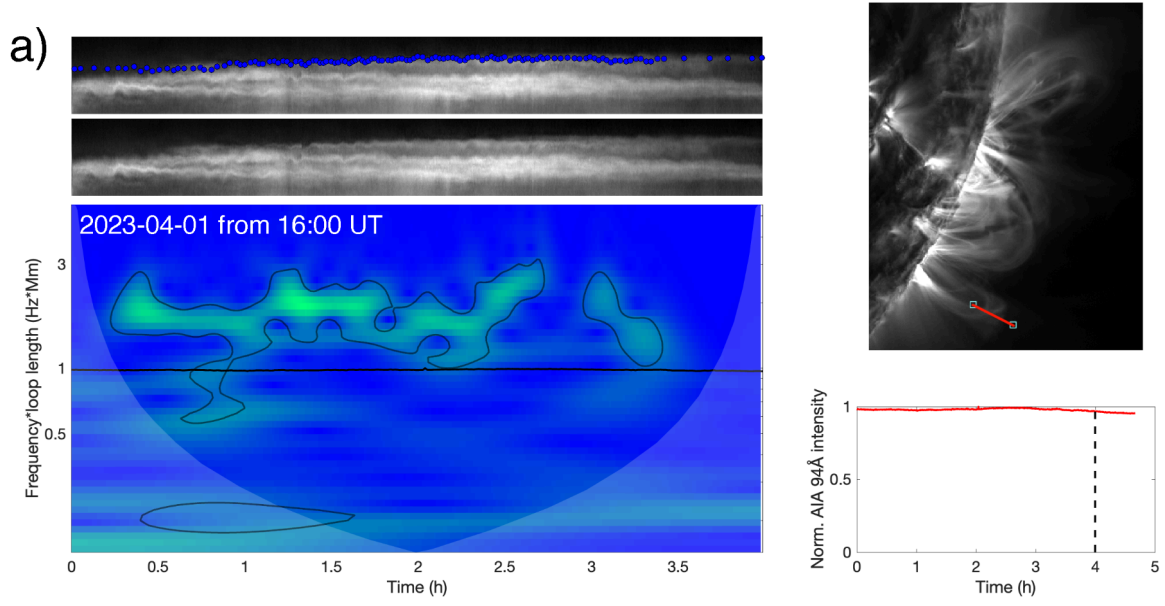

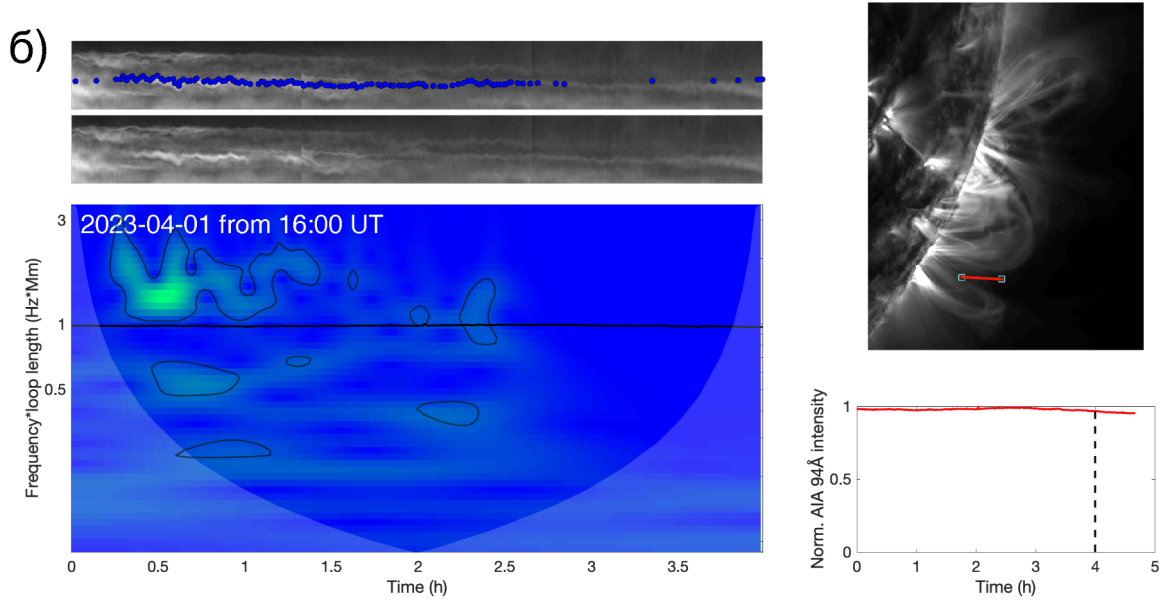

**Fig. 3.** The results of the analysis of two slices for AR 2023-04-01 from 16:00 to 20:00, in which there were no powerful flares similar to those shown in Fig. 1.

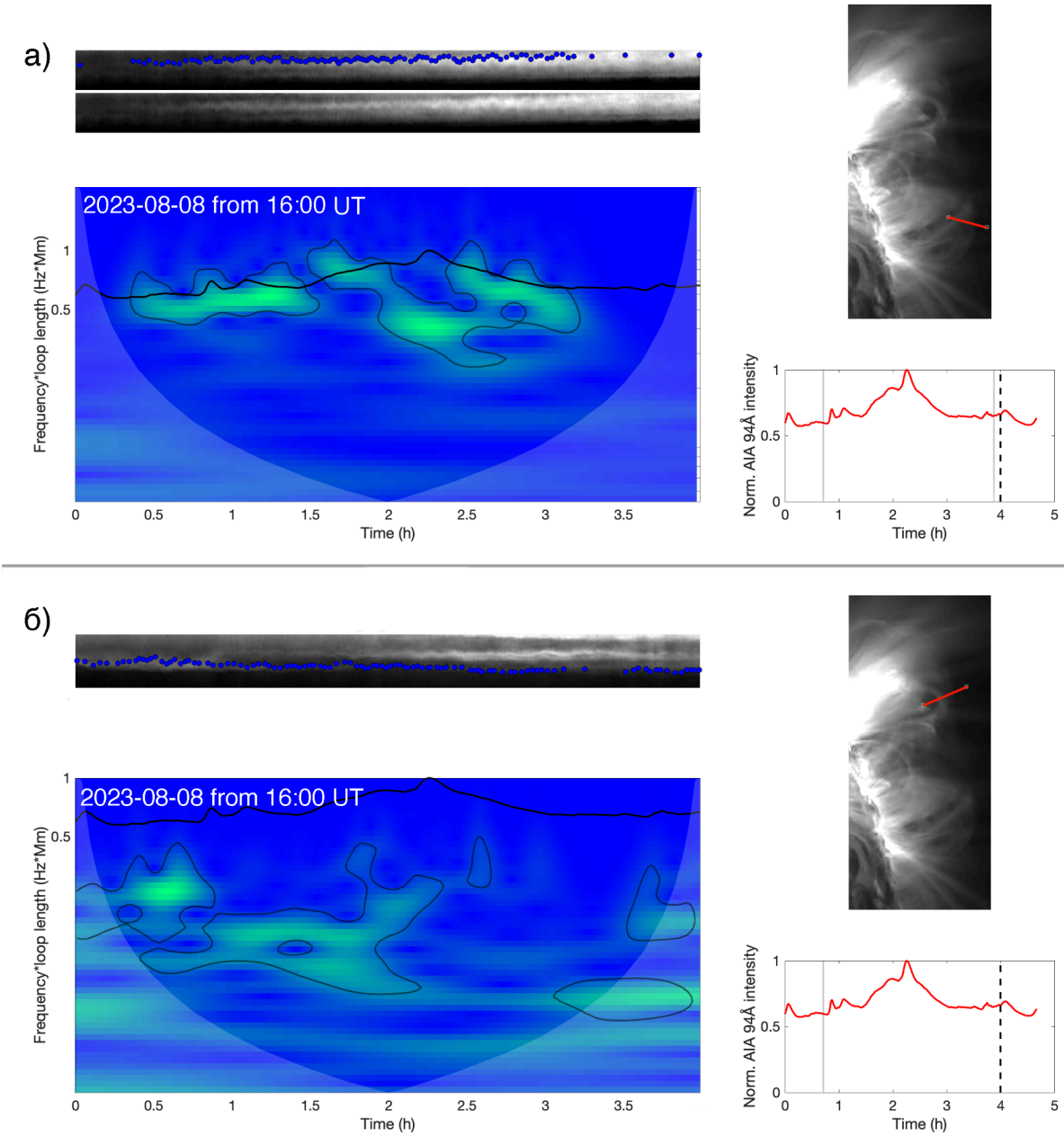

**Fig. 4.** Results of the analysis of two slices for AR 2024-08-08 from 16:00 to 20:00, in which there were no powerful flares similar to those shown in Fig. 1. Weaker flares in the region during the period under review: SOL2023-08-08T16:45 (C1.5), SOL2023-08-08T19:52 (C1.7).

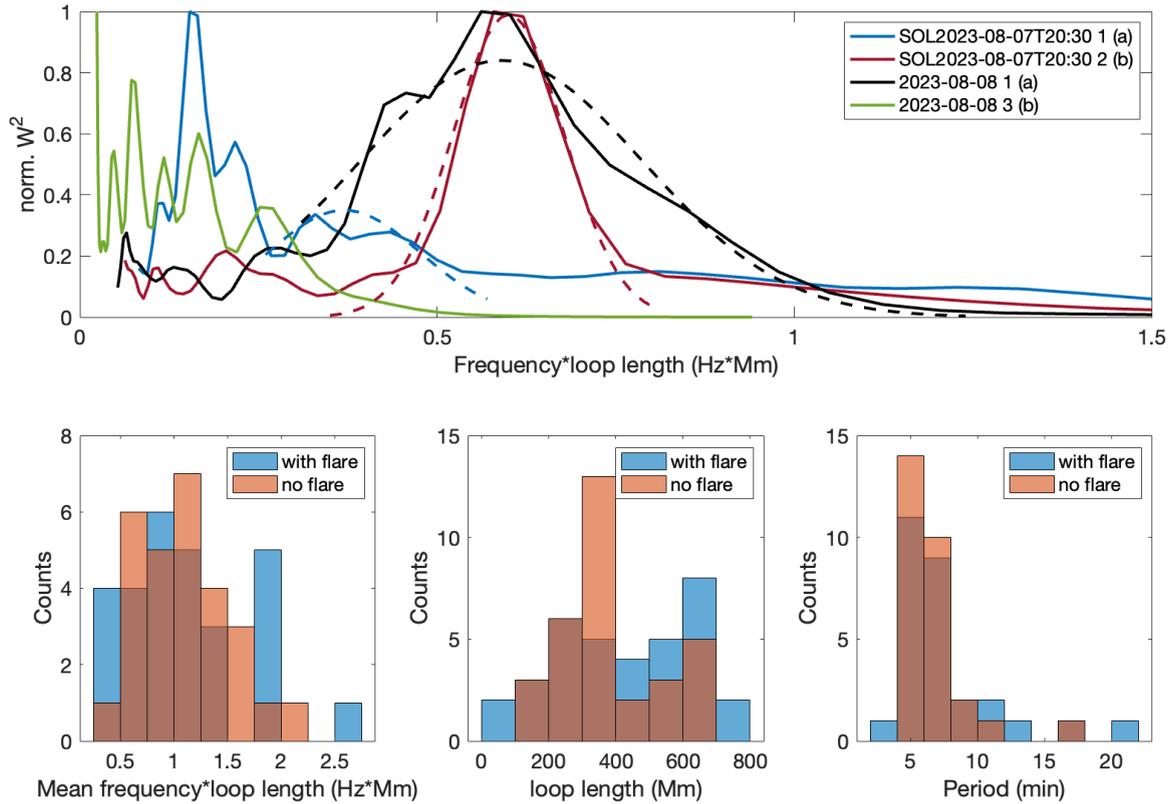

**Fig. 5.** The integral wavelet spectra for the sections shown in Fig. 1a (blue curve) and 1b (red curve) and Fig. 4a (black curve) and 4b (green curve) are shown at the top. The first three spectra are approximated by Gaussians (dotted lines) to find the average frequency multiplied by the loop length. The spectrum for slice 4b cannot be approximated by a Gaussian due to the large number of peaks in it, which is reflected in the table (see Table 2 for 2024-08-08 slice 3). The bottom left shows histograms for the average frequency multiplied by the loop length, estimated from the integral spectra, in blue for ARs with powerful flares and in red for ARs without powerful flares. Similarly, the lower panels in the centre and on the right show histograms for the estimated length of the loops under consideration and the oscillation period, respectively.